# Dielectric Deposition Enhanced Crystallization in Atomic-Layer-Deposited Indium Oxide Transistors Achieving High Gated-Hall Mobility Exceeding 100 $cm^2/V{\cdot}s$ at Room Temperature

*Chen Wang[1], Kai Jiang[1], Jinxiu Zhao[1], Ziheng Wang[1], Guilei Wang[2], Chao Zhao[2], and Mengwei Si[1,*]*

[1]National Key Laboratory of Advanced Micro and Nano Manufacture Technology and School of Information Science and Electronic Engineering, Shanghai Jiao Tong University, Shanghai 200240, China;

[2]Beijing Superstring Academy of Memory Technology, Beijing 100176, China.

*Correspondence authors. Email: mengwei.si@sjtu.edu.cn

## Abstract

In this work, we report high-performance atomic-layer-deposited indium oxide ($In_2O_3$) transistors with high gated-Hall mobility ($\mu_H$) exceeding 100 $cm^2/V\cdot s$ at room temperature (RT). It is found that the deposition of top hafnium oxide ($HfO_2$) above the $In_2O_3$ channel significantly enhances its crystallization, leading to an average grain size of 97.2 nm in a 4.2-nm $In_2O_3$ channel. The ALD of $In_2O_3$ exhibits an epitaxy-like growth behavior, with its (222) planes aligning parallel to the (111) planes of both the top and bottom $HfO_2$ dielectrics. As a result, bottom-gate $In_2O_3$ transistors with a high $\mu_H$ of 100.9 $cm^2/V\cdot s$ and a decent subthreshold swing (SS) of 94 mV/dec are achieved by gated-Hall measurement at RT. Furthermore, the devices maintain excellent performance at low temperatures, achieving a $\mu_H$ of 162.2 $cm^2/V\cdot s$ at 100 K. Our study reveals the critical role of dielectric deposition induced crystallization in enhancing carrier transport and offers a scalable pathway toward high-mobility devices.

Oxide semiconductor transistors have been mature technology in display applications and have gained significant attention as promising candidates for monolithic 3D integration and dynamic random-access memory applications[1–9]. Tremendous efforts have been made to improve the mobility ($\mu$) of oxide semiconductors to enhance the drive current of oxide semiconductor transistors, and high field-effect mobility ($\mu_{FE}$) were reported[10-16]. In oxide semiconductors like amorphous indium gallium zinc oxide (IGZO) and indium oxide ($In_2O_3$), the 5s orbital of indium (In) constitutes the wavefunction of the conduction band bottom and the overlap among the neighboring In 5s orbitals determines the carrier transport path[17]. Therefore, a common method to enhance $\mu_{FE}$ is by increasing the indium composition in amorphous oxides[18-20]. Furthermore, improving the crystallinity of oxides is also an effective way to enhance $\mu_{FE}$, such as annealing-induced crystallization of Ga-doped $In_2O_3$[21], H-doping induced solid-phase crystallization of $In_2O_3$[22], and metal-induced crystallization of $In_2O_3$[23]. However, the aforementioned approaches may suffer from stability challenges due to hydrogen incorporation[24,25], or poor subthreshold swings (SS) possibly due to defect generation, or incorporation of metal elements other than In. Therefore, approaches to enhance the crystallinity of $In_2O_3$ is still needed.

On the other hand, $\mu_{FE}$ can cause the overestimation of intrinsic $\mu$ even in devices with ideal source/drain (S/D) contacts. Therefore, the reported mobility measurements in literatures may not be accurate. In contrast, the Hall mobility ($\mu_H$) can better reflect the intrinsic $\mu$ of oxide semiconductors, in which the conductivity ($\sigma$) and two-dimensional (2D) carrier density ($n_{2D}$) are separately measured. In some of the previous works, the $\mu_H$ of oxide semiconductor thin films (without gate control) was measured[26–28]. However, the $\mu_H$ of thin films cannot reflect the relationship between the $\mu$ and $n_{2D}$ in actual transistors. More importantly, for devices with a threshold voltage ($V_{TH}$) close to 0, the intrinsic $n_{2D}$ of the thin film is relatively low, which usually leads to a low $\mu_H$[22]. At this time, the $\mu_H$ of the thin film cannot accurately represent the intrinsic μ of the device in on-state with high carrier concentration. Therefore, to accurately measure the μ of the semiconductor in a device, the gate-Hall measurement method can be employed to simultaneously measure the $\mu_H$ and $n_{2D}$ at different $V_{GS}$[29]. Currently, to the authors' best knowledge, there haven't been any report of oxide semiconductors exhibiting $\mu_H$ exceeding 100 $cm^2/V{\cdot}s$ by gated-Hall measurements at room temperature (RT)[29–33].

In this work, we have successfully demonstrated high-performance atomic-layer-deposited (ALD) $In_2O_3$ transistors with high gated-Hall mobility ($\mu_H$) exceeding 100 $cm^2/V{\cdot}s$ at RT. This achievement is attributed to the dielectric deposition enhanced crystallization of $In_2O_3$. Especially, the deposition of hafnium oxide ($HfO_2$) on top of the $In_2O_3$ channel significantly enhances the crystallization of $In_2O_3$, resulting in an average grain size of 97.2 nm in a 4.2-nm channel. As a consequence, bottom-gate $In_2O_3$ transistors with an $\mu_H$ of 100.9 $cm^2/V{\cdot}s$ and a decent SS of 94 mV/dec have been achieved by gated-Hall measurements at RT. Additionally, this study highlights the necessity of gated-Hall measurements for the accurate evaluation of the intrinsic μ of oxide semiconductors, as the $\mu_{FE}$ may significantly overestimate the intrinsic μ due to its strong dependence on $V_{GS}$, even in devices with ideal source/drain contacts.

## Results

**Device structures and transfer characteristics of $In_2O_3$ transistors.** Fig. 1(a) shows the schematic diagrams of dual-gate (DG) ALD $In_2O_3$ transistors. Three different device structures including bottom-gate without top $HfO_2$ dielectric deposition (BG w/o top $HfO_2$), bottom-gate with top $HfO_2$ dielectric deposition (BG w/ top $HfO_2$) and top-gate (TG) are also fabricated (Fig. S1). Fig. S2 illustrates the fabrication process flow for the DG transistor, and other structures are fabricated with identical parameters together with certain processes omitted. Figs. 1(b) and 1(c) show the scanning transmission electron microscopy (STEM) images with energy-dispersive X-ray spectroscopy (EDX) mapping at channel region of a DG $In_2O_3$ transistor and a BG $In_2O_3$ transistor without top $HfO_2$, clearly capturing the $HfO_2/In_2O_3/HfO_2$ stack and $HfO_2/In_2O_3$ stack. The gated-Hall bar devices were also fabricated together with the transistors.

Figs. 1(d) and 1(e) present the $I_D$-$V_{GS}$ characteristics of BG $In_2O_3$ transistors with and without top $HfO_2$ with channel length ($L_{ch}$) of 9 μm and at $V_{DS}$ of 0.1 V and 1 V. The corresponding $I_D$-$V_{DS}$ characteristics of the same BG $In_2O_3$ transistor with top $HfO_2$ are plotted in Fig. S3, exhibiting a high maximum $I_D$ of 109 μA/μm and well-behaved drain current saturation at high $V_{DS}$. Figs. 1(f) and 1(g) show the $L_{ch}$-dependent $\mu_{FE}$ and SS of BG $In_2O_3$ transistors with top $HfO_2$ measured at 300 K and 100 K, and BG $In_2O_3$ transistors without top $HfO_2$ measured at 300 K. When $L_{ch}$ is less than 9 μm, the $\mu_{FE}$ decreases with $L_{ch}$, indicating that the $\mu_{FE}$ is underestimated due to the presence of contact resistance. Therefore, when $L_{ch}$ is greater than 9 μm, the $\mu_{FE}$ is relatively reliable. $\mu_{FE}$ versus $V_{GS}$ characteristics of $In_2O_3$ transistors with various structures are plotted in Fig. S4. $\mu_{FE}$ of BG $In_2O_3$ transistors with top $HfO_2$ reaches 124.8 $cm^2/V{\cdot}s$ at 300 K and 137.9 $cm^2/V{\cdot}s$ at 100 K, which is much higher than that of BG $In_2O_3$ transistors without top $HfO_2$. Figs. 1(h) and 1(i) plot the $I_D$-$V_{GS}$ characteristics of TG and DG $In_2O_3$ transistors with $L_{ch}$ of 9 μm at $V_{DS}$ of 0.1 V and 1 V at 300 K. A large counterclockwise hysteresis exists in both devices due to defect generation at top $HfO_2/In_2O_3$ interface, as previously reported[34]. As a result, additional carriers are generated at high $V_{GS}$ to maintain the electrical neutrality in channel. Therefore, $I_D$ of TG and DG transistors are higher than that of BG transistors. However, $\mu_{FE}$ cannot be accurately extracted from $I_D$-$V_{GS}$ characteristics of TG and DG transistors due to the existence of hysteresis, as shown in Figs. S4(c) and S4(d). The SS of TG and DG transistors extracted from forward-sweep transfer curves are 77 mV/dec and 66 mV/dec, and that extracted from reverse-sweep transfer curves are 36 mV/dec and 35 mV/dec, respectively. The SS of TG and DG that exceed the Boltzmann limit are also caused by hysteresis.

**Hall measurement of $In_2O_3$ transistors.** It is well known that $\mu_H$ can more accurately reflect the intrinsic $\mu$ of the $In_2O_3$ channel, because $\sigma$ and $n_{2D}$ are measured separately. False-colored optical microscope image of a DG $In_2O_3$ gated-Hall bar device is shown in Fig. 2(a). The non-overlap between S/D and gate electrodes excludes the influence of S/D contacts on Hall measurements. The voltage drop parallel to the channel is determined by measuring the voltage difference at two points labeled $V_{XX}$, and the Hall voltage is determined by measuring the voltage difference at two points labeled $V_{XY}$. Figs. 2(b), 2(c) and 2(d) present the $I_D$, voltage parallel to $I_D$ ($V_{XX}$) and voltage perpendicular to $I_D$ ($V_{XY}$) versus $V_{GS}$ characteristics of the BG $In_2O_3$ Hall bar device with top $HfO_2$ at 300 K with $V_{DS}$ of 0.1 V and magnetic field from -10 T to 10 T. $I_D$ of the Hall bar shows a similar trend to the transistor and is nearly unaffected by magnetic field. $V_{XX}$ decreases at high $V_{GS}$ as a result of contact resistance. $\sigma$ is calculated based on the $I_D$ and $V_{XX}$. Due to the Hall bar structure accurately measuring the voltage drop of the channel, the $\mu_{FE}$ can be accurately measured, as shown in Fig. 2(e). $\mu_{FE}$ decreases at high temperature, and the maximum of $\mu_{FE}$ at 300 K is 126.7 $cm^2/V \cdot s$, which is consistent with the $\mu_{FE}$ of transistor. $V_{XY}$ varies with external magnetic field, from where the $n_{2D}$ is evaluated. Fig. 2(f) plots the $\mu_H$ versus $V_{GS}$ characteristics of the BG $In_2O_3$ Hall bar device with top $HfO_2$ measured from 4 K to 300 K. At small $V_{GS}$ ($V_{GS}$ <1 V) and low carrier density, $\mu_H$ increases with $V_{GS}$, which is likely caused by percolation mechanism and trap-limited conduction[35,36]. The $\mu_H$ at small $V_{GS}$ can be described as a power law, $\mu_H = \mu_{L0}(n_{2D}/n_{L0})^{\gamma}$, where $\mu_{L0}$ is the constant mobility parameter, $n_{L0}$ is the critical carrier concentration, and $\gamma$ is the model parameter. At large $V_{GS}$ ($V_{GS}$ >1 V) and high carrier density $n_{2D}$, $\mu_H$ decreases with $n_{2D}$ because of the high field scattering due to surface roughness and phonon scattering[37], which can be described as $\mu_H = \mu_{H0}(n_{2D}/n_{H0})^{-r}$, where $\mu_{H0}$ is the constant mobility parameter, $n_{H0}$ is the critical carrier concentration, and r is 2 for surface roughness scattering and 0.3 for phonon scattering. The $\mu_H$ reaches 100.9 $cm^2/V{\cdot}s$ at 300 K and increases at low temperature due to the suppressed phonon scattering. Fig. 2(g) plots the $n_{2D}$ versus $V_{GS}$ characteristics of the BG $In_2O_3$ Hall bar device with top $HfO_2$ from 4 K to 300 K, which satisfies the relationship of $n_{2D}=C_{OX}(V_{GS}-V_{TH})/q$, where $C_{OX}$ is the gate capacitance and q is elementary charge, suggesting a low trap density in bulk $In_2O_3$ and at the interface with $HfO_2$ when Fermi level ($E_F$) aligns at the on-state of the device. The $C_{OX}$ is determined by measuring capacitance-voltage (CV) characteristics of Mo/$HfO_2$/Mo stack as shown in Fig. S5, which is determined to be 1.31 $\mu F/cm^2$.

Figs. 2(h) and 2(i) plot the $\mu_H$ and $n_{2D}$ versus $V_{GS}$ characteristics measured at 300 K from $In_2O_3$ Hall bar devices using BG with and without top $HfO_2$, TG and DG structures. The BG device without top $HfO_2$ shows a relatively low $\mu_H$ of 75.3 $cm^2/V{\cdot}s$ compared with other devices. $\mu_H$ of BG with top $HfO_2$, TG and DG devices are similar and higher than that of the BG device without top $HfO_2$, suggesting $\mu_H$ is improved by the top $HfO_2$ dielectric deposition. Among them, TG device achieves the highest $\mu_H$ of 109.2 $cm^2/V{\cdot}s$ due to a higher $n_{2D}$, as shown in Fig. 2(i). The higher $n_{2D}$ of TG device confirms the generation of defects with positive charge at top $HfO_2/In_2O_3$ interface.

**$\mu_H$ and $\mu_{FE}$ of $In_2O_3$ transistors.** Temperature-dependent $\mu_H$ of the four different devices are plotted in Fig. 3(a). A high $\mu_H$ of 162.2 $cm^2/V{\cdot}s$ at 100 K is achieved on a DG device. The $\mu_{FE}$ is further extracted from a BG $In_2O_3$ Hall bar device with top $HfO_2$. The impact of contact resistance can be minimized because the $L_{CH}$ of the Hall bar device is rather large and $V_{GS}$-dependent contact resistance can also be excluded due to the non-overlapping gate electrodes and S/D electrodes. The relationship between $\mu_{FE}$ and intrinsic μ is achieved as by combining eqn. (S1) and (S2)[38],

$$\mu_{FE} = \frac{\partial \mu}{\partial V_{GS}}\left(V_{GS} - V_{TH} - \frac{V_{DS}}{2}\right) + \mu \quad (1)$$

A deviation between $\mu_{FE}$ and intrinsic μ arises from the dependence of intrinsic μ on $V_{GS}$[38]. In oxide semiconductors, the dependence of intrinsic μ on $V_{GS}$ cannot be ignored due to the multiple scattering mechanisms and the percolation conduction. In contrast, $\mu_H$ can better represent the intrinsic μ because the $\mu_H$ is derived directly by carrier density and conductivity according to σ=qnμ, which aligns with the physical definition of μ[39]. Fig. S6 compares $\mu_H$, experimental $\mu_{FE}$, and calculated $\mu_{FE}$ from $\mu_H$ by eqn. (1) of a BG $In_2O_3$ Hall bar device with top $HfO_2$ at 300 K. The calculated $\mu_{FE}$ from $\mu_H$ matches well with experimental $\mu_{FE}$, which confirms the above analysis. The $\mu_{FE}$ is much higher than that of $\mu_H$ because of the large positive $\partial\mu/\partial V_{GS}$. Similarly, $\mu_{FE}$ decreases significantly at high $V_{GS}$ because of the large negative $\partial\mu/\partial V_{GS}$. Fig. 3(b) shows the temperature-dependent $\mu_{FE}$ from 4 K to 300 K of BG $In_2O_3$ Hall bar devices with and without top $HfO_2$. A $\mu_{FE}$ of 142.4 $cm^2/V{\cdot}s$ at 300 K is achieved on the BG device with top $HfO_2$, which is further enhanced to 192.5 $cm^2/V{\cdot}s$ at 100 K.

The $\mu_H$ versus SS of the $In_2O_3$ transistors in this work are benchmarked with state-of-the-art oxide semiconductors transistors with gated-Hall measurements (noted as gated Hall) and Hall measurement on oxide semiconductor film without gate (noted as film Hall)[22,29,30,32,33,40–42], as shown in Fig. 3(c). High $\mu_H$ of 162.2 $cm^2/V{\cdot}s$ and SS of 61 mV/dec at 100 K are achieved for DG device, and $\mu_H$ of 100.9 $cm^2/V{\cdot}s$ and SS of 94 mV/dec at 300 K are achieved for BG device with top $HfO_2$. The simultaneous achievement of high $\mu_H$ and steep SS confirms the effectiveness of the proposed dielectric deposition enhanced crystallization of $In_2O_3$ for high mobility devices, as will discussed in next section.

**Structural properties of the $HfO_2/In_2O_3/HfO_2$ stack and $HfO_2/In_2O_3$ stack.** According to the above results, the deposition of top $HfO_2$ dielectrics enables a significantly higher $\mu_H$. To further investigate the mechanism of mobility enhancement, Figs. 4(a) and 4(b) show the high-resolution TEM (HRTEM) cross-sectional images of $HfO_2/In_2O_3/HfO_2$ stack and $HfO_2/In_2O_3$ stack. In the $HfO_2/In_2O_3/HfO_2$ stack, distinct lattice fringes are observed within the $In_2O_3$ layer, whereas in the $HfO_2/In_2O_3$ stack, no such clear lattice fringes are presented in the $In_2O_3$ layer. Hence, the deposition of top $HfO_2$ dielectric enhances the crystallization of $In_2O_3$. The (111) planes of upper and lower $HfO_2$ and the (222) plane of $In_2O_3$ align in the same direction, exhibiting a behavior similar to epitaxial growth, suggesting the crystallization of $In_2O_3$ is induced by both top and bottom $HfO_2$ deposition. Additional HRTEM images of $HfO_2/In_2O_3/HfO_2$ stack and $HfO_2/In_2O_3$ stack are given in Fig. S7, which further supports the above statements. Fig. 4(c) plots grazing-incidence X-ray diffraction (GIXRD) spectra of $HfO_2/In_2O_3/HfO_2$ stack, $HfO_2/In_2O_3$ stack, and the single $HfO_2$ layer together with standard cubic $In_2O_3$ (c-$In_2O_3$) and monoclinic $HfO_2$ (m-$In_2O_3$) patterns. The $In_2O_3$ in $HfO_2/In_2O_3/HfO_2$ stack has a higher diffraction intensity compared with that in $HfO_2/In_2O_3$ stack, confirming its higher crystallization. The XRD spectrum of the single $HfO_2$ layer matches that of m-$HfO_2$, indicating that the phase structure of $HfO_2$ is the monoclinic phase, which is consistent with the HRTEM results. Figs. 4(d) and 4(e) present the electron backscattered diffraction (EBSD) images of $HfO_2/In_2O_3/HfO_2$ stack and $HfO_2/In_2O_3$ stack. Fig. 4(f) shows the distribution of grain size extracted from EBSD data. The average grain sizes of $In_2O_3$ in $HfO_2/In_2O_3/HfO_2$ stack and $HfO_2/In_2O_3$ stack are 97.2 nm and 51.7 nm, respectively, further validating the enhanced crystallization by the top $HfO_2$ dielectric deposition, which well explains the origin of mobility enhancement due to enhanced crystallinity.

## Discussion

This study presents the demonstration of high-performance ALD $In_2O_3$ transistors with high $\mu_H$ exceeding 100 $cm^2/V\cdot s$ together with a decent SS of 94 mV/dec at room temperature. The $\mu_H$ is further enhanced to 162.2 $cm^2/V\cdot s$ at 100 K due to the suppression of phonon scattering. This enhancement is primarily attributed to the dielectric deposition induced crystallization of the $In_2O_3$ channel. The introduction of top and bottom $HfO_2$ dielectric leads to epitaxy-like growth behavior, aligning the (222) planes of $In_2O_3$ with the (111) planes of $HfO_2$. This process increases the average grain size from 51.7 nm to 97.2 nm, thereby facilitating more efficient carrier transport. These findings provide a practical and scalable approach to improve mobility through crystallinity engineering. Besides, the gated-Hall measurement technique employed in this study provides a more accurate assessment of the intrinsic $\mu$, avoiding the overestimation of mobility by $\mu_{FE}$.

## Methods

**Fabrication of $In_2O_3$ transistors.** The fabrication process started with ALD deposition of 40-nm $Al_2O_3$ on Si substrate for insulation. 25-nm Mo was deposited by DC magnetron sputtering and patterned by lift-off process as BG electrode. A low sputtering pressure (~0.01 Pa) was used to ensure low surface roughness of the Mo electrode. 8-nm $HfO_2$ bottom dielectric was grown by ALD at 200 °C with Tetrakis(dimethylamido)hafnium (TDMAHf) and $H_2O$ as precursors of Hf and O, respectively. $In_2O_3$ channel was grown by ALD at 225 °C with [3-(dimethylamino)propyl] dimethyl indium (DADI) and ozone ($O_3$) as precursors of In and O, respectively. The channel region was isolated by concentrated hydrochloric acid wet etching. Post semiconductor deposition annealing was carried out in $O_2$ at 350 °C for 5 min. 80-nm Mo was deposited by DC magnetron sputtering as S/D electrodes with the same parameters as BG electrodes. 8-nm $HfO_2$ was grown by ALD at 200 °C as top gate dielectric with the same parameters as bottom dielectric and followed by post dielectric deposition annealing in $O_2$ at 350 °C for 5 min. Finally, 30-nm Ni was thermally evaporated as TG electrode.

**Characterization of the $In_2O_3$ transistors.** The $In_2O_3$ film and transistor structures were evaluated through high-resolution and scanning transmission electron microscopy (HRTEM and STEM) with energy-dispersive X-ray spectroscopy (EDX) (Thermo Scientific, Talos F200X G2), grazing incidence X-ray diffraction (GIXRD) (Rigaku, SmartLab) with CuKα radiation and electron backscattering diffraction (EBSD) (Zeiss, Sigma 500). The IV characteristics were measured using a semiconductor parameter analyzer (Keysight, B1500A) in the vacuum and dark. The gated Hall bar devices were fabricated together with TFTs to evaluate the Hall mobility ($\mu_H$) and carrier density ($n_{2D}$). The Hall measurement was performed based on the ASTM F76 standard with a Hall bar structure to avoid the distortion of geometry in the regularly used van der Pauw samples. The $V_{XX}$ and $V_{XY}$ were measured by a lock-in amplifier (Stanford Research Systems, SR830), and the $V_{GS}$ was applied by source-meter unit (Keithley, 2400). The gated Hall measurement was carried out in a physical property measurement system (PPMS) (Quantum Design, DynaCool-14T). The driven AC signal of $V_{DS}$ has a frequency of 17.7 Hz and an amplitude of 0.1 $V_{RMS}$.

REFERENCE

1. Nathan, A. & Jeon, S. Oxide Electronics: Translating Materials Science from Lab-to-Fab. *MRS Bull.* **46**, 1028–1036 (2021).
2. Lu, W. *et al.* First Demonstration of Dual-Gate IGZO 2T0C DRAM with Novel Read Operation, One Bit Line in Single Cell, $I_{ON}$=1500μA/μm@$V_{DS}$=1V and Retention Time>300s. In *2022 International Electron Devices Meeting* 611-614 (2022).
3. Geng, D. *et al.* Thin-film Transistors for Large-Area Electronics. *Nat. Electron.* **6**, 963–972 (2023).
4. Kim, W. *et al.* Demonstration of Crystalline IGZO Transistor with High Thermal Stability for Memory Applications. In *2023 Symposium on VLSI Technology* T17-4 (2023).
5. Datta, S. *et al.* Amorphous Oxide Semiconductors for Monolithic 3D Integrated Circuits. In *2024 Symposium on VLSI Technology* TFS1 (2024).
6. Sarkar, E. *et al.* First Demonstration of W-doped $In_2O_3$ Gate-All-Around (GAA) Nanosheet FET with Improved Performance and Record Threshold Voltage Stability. In *2024 International Electron Devices Meeting* 12–1 (2024).
7. Seo, D. *et al.* Transport Properties of Crystalline IGZO Channel Devices: Effects of Cation Disorders, Composition and Dimensions. In *2024 International Electron Devices Meeting* 32–5 (2024).
8. Yang, J. E. *et al.* A-IGZO FETs with High Current and Remarkable Stability for Vertical Channel Transistor (VCT) / 3D DRAM Applications. In *2024 Symposium on VLSI Technology* T4.5 (2024).

9. Tang, B. *et al.* First Demonstration of Fluorine-Treated IGZO FETs with Record-Low Positive Bias Temperature Instability ($|\Delta V_{TH}|$ < 44 mV) at an Elevated Temperature (395 K). In *2025 Symposium on VLSI Technology* T17-2 (2025).

10. Shiah, Y. S. *et al.* Mobility–stability trade-off in oxide thin-film transistors. *Nat. Electron.* **4**, 800–807 (2021).

11. Chiang, K. H. *et al.* Integration of 0.75 V $V_{DD}$ Oxide-Semiconductor 1T1C Memory with Advanced Logic for an Ultra-Low-Power Low-Latency Cache Solution. In *2025 Symposium on VLSI Technology* T2-1 (2025).

12. Cho, M. H. *et al.* High-Performance Amorphous Indium Gallium Zinc Oxide Thin-Film Transistors Fabricated by Atomic Layer Deposition. *IEEE Electron Device Lett.* **39**, 688–691 (2018).

13. Si, M. *et al.* Scaled Indium Oxide Transistors Fabricated Using Atomic Layer Deposition. *Nat. Electron.* **5**, 164–170 (2022).

14. Hikake, K. *et al.* A Nanosheet Oxide Semiconductor FET Using ALD InGaOx Channel and InSnOx Electrode with Normally-off Operation, High Mobility and Reliability for 3D Integrated Devices. In *2023 Symposium on VLSI Technology* T14-1 (2023).

15. Kang, Y. *et al.* Thickness-Engineered Extremely-thin Channel High Performance ITO TFTs with Raised S/D Architecture: Record-Low $R_{SD}$, Highest Moblity (Sub-4 nm $T_{CH}$ Regime), and High $V_{TH}$ Tunability. In *2023 Symposium on VLSI Technology* T11-2 (2023).

16. Kim, T. *et al.* Ultrahigh Field-Effect Mobility of 147.5 $cm^2/Vs$ in Ultrathin $In_2O_3$ Transistors via Passivating the Surface of Polycrystalline $HfO_2$ Gate Dielectrics. *Appl. Phys. Lett.* **126**, 033501 (2025).

17. Nomura, K. *et al.* Room-Temperature Fabrication of Transparent Flexible Thin-Film Transistors Using Amorphous Oxide Semiconductors. *Nature* **432**, 488–492 (2004).

18. Sheng, J. *et al.* Amorphous IGZO TFT with High Mobility of 70 $cm^2/(V\ s)$ via Vertical Dimension Control Using PEALD. *ACS Appl. Mater. Interfaces* **11**, 40300–40309 (2019).

19. Han, K. *et al.* Indium-Tin-Oxide Thin-Film Transistors With High Field-Effect Mobility (129.5 $cm^2/V \cdot s$) and Low Thermal Budget (150 °C). *IEEE Electron Device Lett.* **44**, 1999–2002 (2023).

20. Charnas, A. *et al.* Review—Extremely Thin Amorphous Indium Oxide Transistors. *Adv. Mater.* **36**, 2304044 (2024).

21. Takahashi, T. *et al.* ALD Polycrystalline Ga-Doped $In_2O_3$ (Poly-IGO) Nanosheet Exceeding Intrinsic Mobility of 120 $cm^2/Vs$ for Process-Friendly BEOL-Compatible FET Application. In *2025 Symposium on VLSI Technology* T12-2 (2025).

22. Magari, Y. *et al.* High-Mobility Hydrogenated Polycrystalline $In_2O_3$ ($In_2O_3$:H) Thin-Film Transistors. *Nat. Commun.* **13**, 1078 (2022).

23. Chen, Z. *et al.* High-Crystallinity and Enhanced Mobility in $In_2O_3$ Thin-Filam Transistors via Metal-Induced Method. *Appl. Phys. Lett.* **126**, 033504 (2025).

24. Lin, Z. *et al.* The Role of Oxygen Vacancy and Hydrogen on the PBTI Reliability of ALD IGZO Transistors and Process Optimization. *IEEE Trans. Electron Devices* **71**, 3002–3008 (2024).

25. Lin, Z. *et al.* The Critical Role of Passivation Layer and Semiconductor Interface on the Hydrogen Stability of ALD IGZO Transistors. *IEEE Trans. Electron Devices* **72**, 4138–4142 (2025).

26. Newhouse, P. F. *et al.* High Electron Mobility W-doped $In_2O_3$ Thin Films by Pulsed Laser Deposition. *Appl. Phys. Lett.* **87**, 112108 (2005).

27. Koida, T. & Kondo, M. High-Mobility Transparent Conductive Zr-doped $In_2O_3$. *Appl. Phys. Lett.* **89**, 082104 (2006).

28. Bierwagen, O. & Speck, J. S. High Electron Mobility $In_2O_3$ (001) and (111) Thin Films with Nondegenerate Electron Concentration. *Appl. Phys. Lett.* **97**, 072103 (2010).

29. Socratous, J. *et al.* Energy-Dependent Relaxation Time in Quaternary Amorphous Oxide Semiconductors Probed by Gated Hall Effect Measurements. *Phys. Rev. B* **95**, 045208 (2017).

30. Li, S. *et al.* Nanometre-thin Indium Tin Oxide for Advanced High-performance Electronics. *Nat. Mater.* **18**, 1091–1097 (2019).

31. Anders, J. *et al.* Gated Hall and Field-Effect Transport Characterization of E-mode ZnO TFTs. *Appl. Phys. Lett.* **116**, 252105 (2020).

32. Imanishi, K., Matsuda, T. & Kimura, M. Analysis of Carrier Mobility in Amorphous Metal-Oxide Semiconductor Thin-Film Transistor Using Hall Effect. *IEEE Electron Device Lett.* **41**, 1025–1028 (2020).

33. Hu, K. *et al.* Tri-Layer Heterostructure Channel of a-IGZO/a-ITZO/a-IGZO Toward Enhancement of Transport and Reliability in Amorphous Oxide Semiconductor Thin Film Transistors. *Adv. Electron. Mater.* **11**, 2400266 (2024).

34. Jiang, K. *et al.* Top-Gate Atomic-Layer-Deposited Oxide Semiconductor Transistors With Large Memory Window and Non-Ferroelectric $HfO_2$ Gate Stack. *IEEE Electron Device Lett.* **46**, 1353–1356 (2025).

35. Kamiya, T., Nomura, K. & Hosono, H. Electronic Structures Above Mobility Edges in Crystalline and Amorphous In-Ga-Zn-O: Percolation Conduction Examined by Analytical Model. *J. Disp. Technol.* **5**, 462–467 (2009).

36. Lee, S. *et al.* Trap-Limited and Percolation Conduction Mechanisms in Amorphous Oxide Semiconductor Thin Film Transistors. *Appl. Phys. Lett.* **98**, 203508 (2011).

37. Kim, M. J. *et al.* Effect of Channel Thickness on Performance of Ultra-Thin Body IGZO Field-Effect Transistors. *IEEE Trans. Electron Devices* **69**, 2409–2416 (2022).

38. Wang, C. *et al.* On the Accurate Evaluation of Intrinsic Electron Mobility on Oxide Semiconductor Transistors. *IEEE Electron Device Lett.* (2025).

39. Schroder, D. K. *Semiconductor Material and Device Characterization*. *John Wiley & Sons* (2015).

40. Hwang, A. Y. *et al.* Metal-Induced Crystallization of Amorphous Zinc Tin Oxide Semiconductors for High Mobility Thin-Film Transistors. *Appl. Phys. Lett.* **108**, 152111 (2016).

41. Xu, H. *et al.* Improvement of Mobility and Stability in Oxide Thin-Film Transistors Using Triple-Stacked Structure. *IEEE Electron Device Lett.* **37**, 57–59 (2016).

42. Jeong, S. G., Jeong, H. J. & Park, J. S. Low Subthreshold Swing and High Performance of Ultrathin PEALD InGaZnO Thin-Film Transistors. *IEEE Trans. Electron Devices* **68**, 1670–1675 (2021).

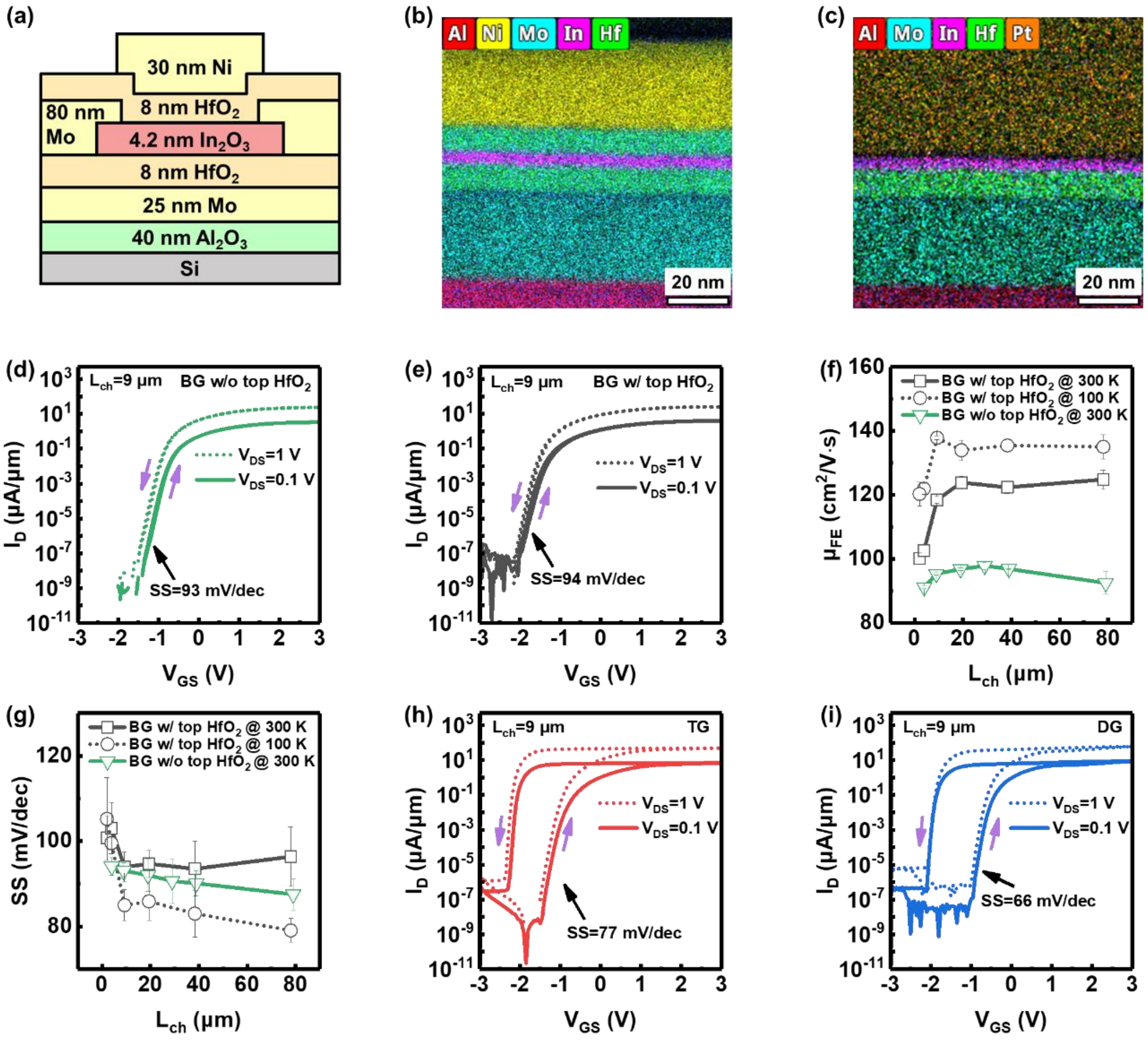

Figure 1. **Device structures and transfer characteristics of $In_2O_3$ transistors.** (a) Schematic diagrams of ALD dual-gate $In_2O_3$ transistors. STEM cross-sectional image of $In_2O_3$ transistors with various device structures with EDX elemental mapping, (b) DG and (c) BG w/o top $HfO_2$. $I_D$-$V_{GS}$ characteristics of BG $In_2O_3$ transistors (d) with and (e) without top $HfO_2$ with $L_{ch}$ of 9 µm and at $V_{DS}$ of 0.1 V at 300 K. $L_{ch}$-dependent (f) $\mu_{FE}$ and (g) SS of BG $In_2O_3$ transistors with top $HfO_2$ measured at 300 K and 100 K, and BG $In_2O_3$ transistor without top $HfO_2$ measured at 300 K. $I_D$-$V_{GS}$ characteristics of (h) TG and (i) DG $In_2O_3$ transistors with $L_{ch}$ of 9 µm and at $V_{DS}$ of 0.1 V at 300 K. The SS are extracted from forward-sweep transfer curves.

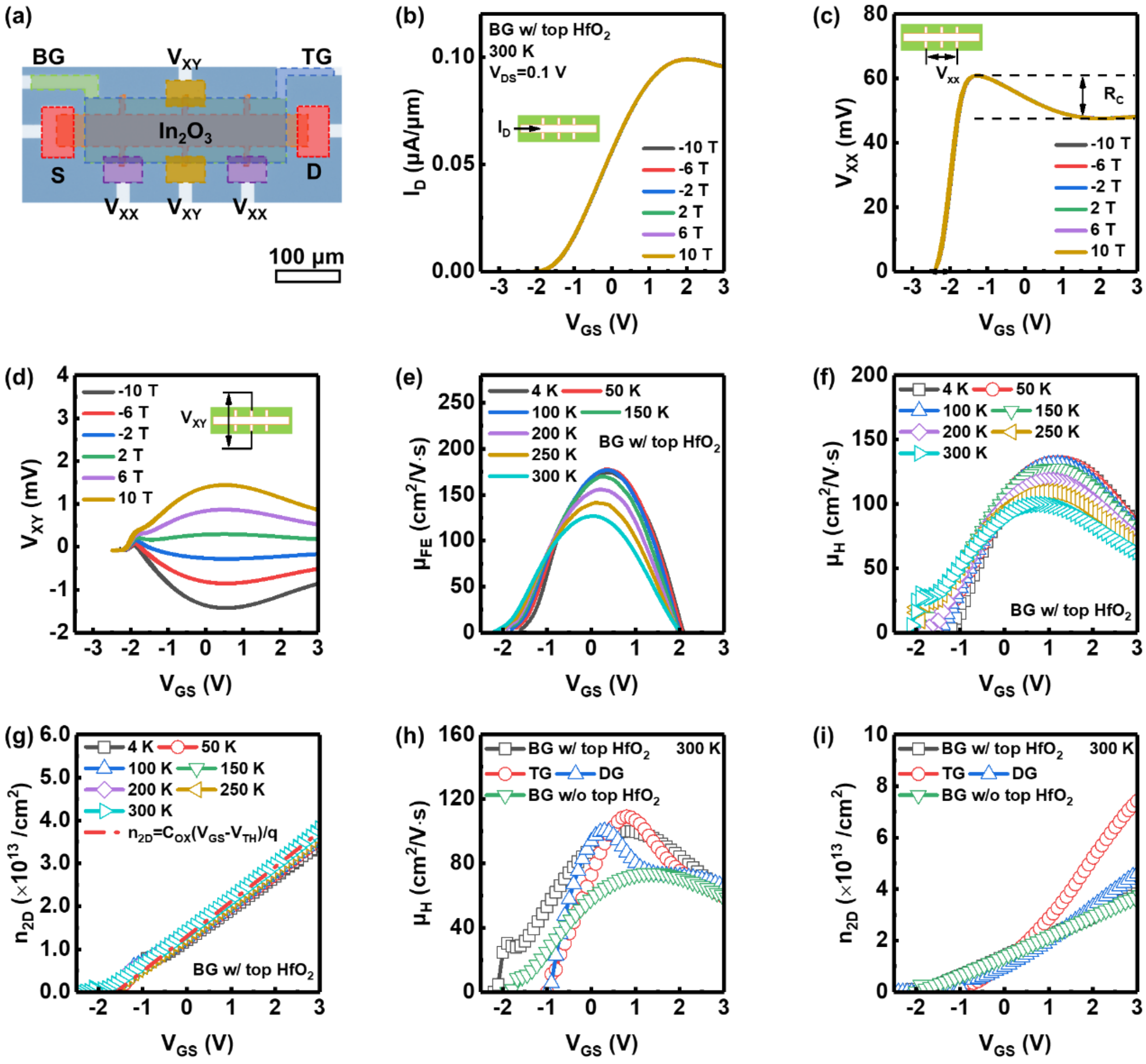

Figure 2. **Hall measurement of $In_2O_3$ transistors.** (a) False-colored optical microscope image of a DG $In_2O_3$ gated-Hall bar device. (b) $I_D$, (c) $V_{XX}$, and (d) $V_{XY}$ versus $V_{GS}$ characteristics of a BG $In_2O_3$ Hall bar with top $HfO_2$ measured at 300 K and with magnetic field from -10 T to 10 T. The $V_{DS}$ is 0.1 V. (e) $\mu_{FE}$ versus $V_{GS}$ characteristics of a BG $In_2O_3$ Hall bar with top $HfO_2$ avoiding the influence of contact resistance. (f) $\mu_H$ and (g) $n_{2D}$ versus $V_{GS}$ characteristics of a BG $In_2O_3$ Hall bar with top $HfO_2$ measured from 4 K to 300 K. $n_{2D}$ determined by $n_{2D}=C_{OX}(V_{GS}-V_{TH})/q$ is plotted by the dash line in (g). The $C_{OX}$ is determined by measuring CV of Mo/$HfO_2$/Mo stack, which is 1.31 μF/$cm^2$. (h) $\mu_H$ and (i) $n_{2D}$ versus $V_{GS}$ characteristics measured at 300 K from Hall bar devices using BG with and without top $HfO_2$, TG and DG structures.

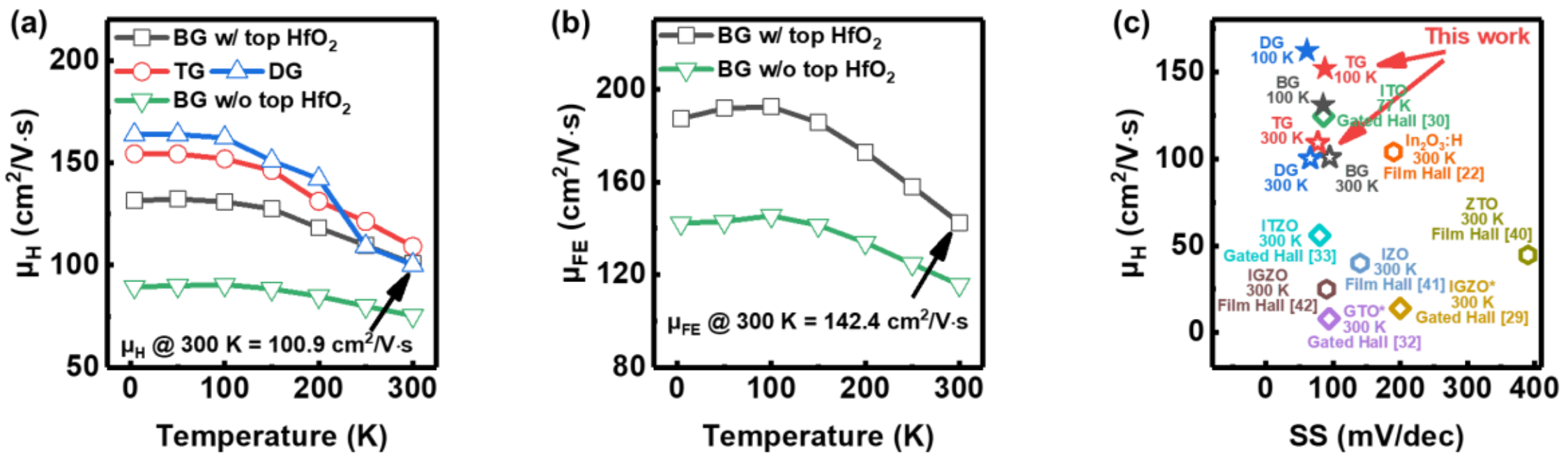

Figure 3. **$\mu_H$ and $\mu_{FE}$ of $In_2O_3$ transistors.** Temperature-dependent (a) $\mu_H$ of devices with BG with and without top $HfO_2$, TG, and DG structures, (b) $\mu_{FE}$ of BG $In_2O_3$ Hall bar devices with and without top $HfO_2$ measured from 4 K to 300 K. (c) Benchmarking on the $\mu_H$ versus SS characteristics of $In_2O_3$ devices in this work and oxide semiconductor devices in other reports, including gated-Hall and film Hall measurements. For data point marked with '*', SS is calculated from the $I_D$-$V_{GS}$ curves in the literatures. For film Hall measurements, the $\mu_H$ and SS are not achieved on the same device.

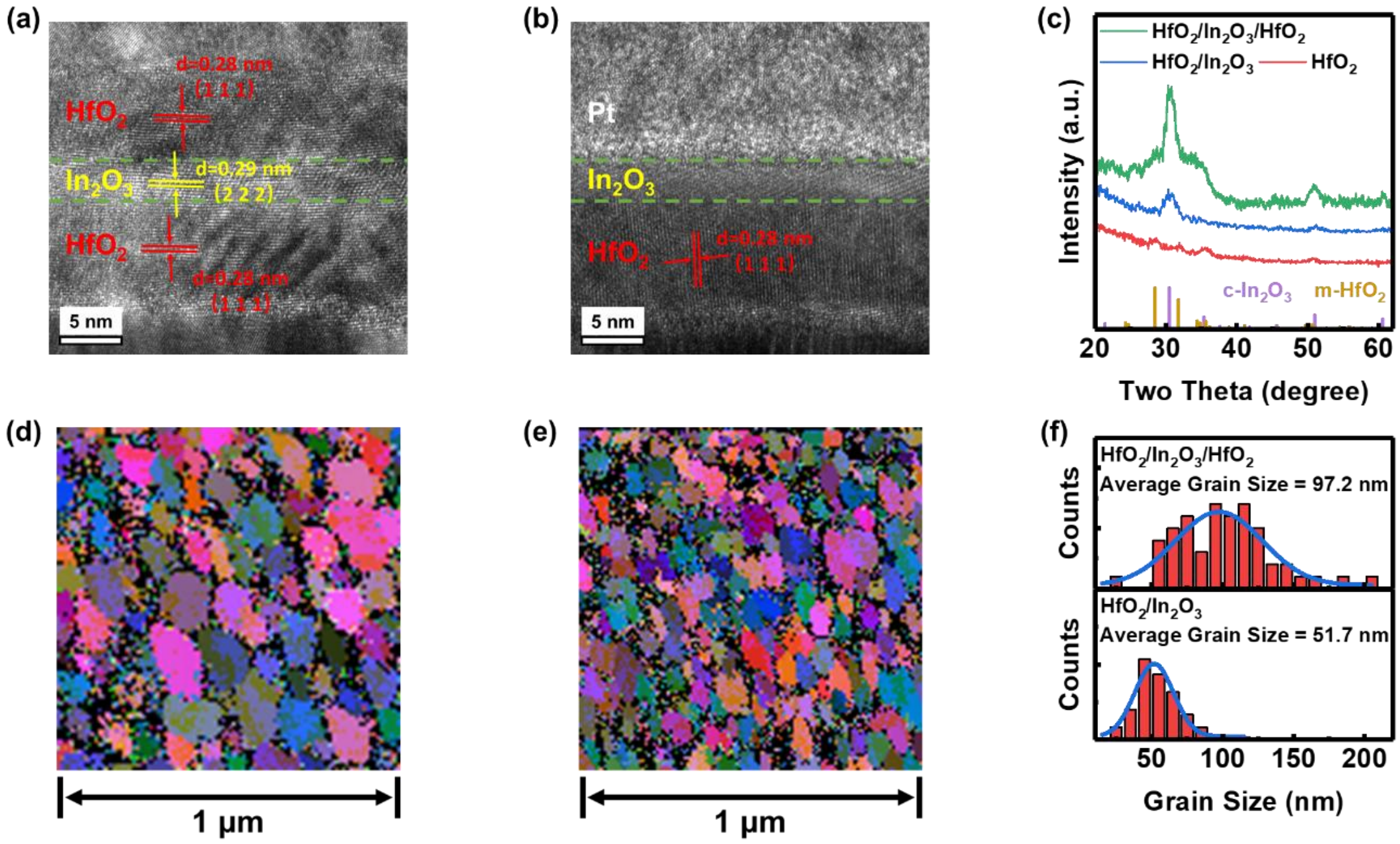

Figure 4. **Structural properties of the $HfO_2/In_2O_3/HfO_2$ stack and $HfO_2/In_2O_3$ stack.** HRTEM cross-section images of (a) $HfO_2/In_2O_3/HfO_2$ stack and (b) $HfO_2/In_2O_3$ stack. The top $HfO_2$ dielectric deposition enhances the crystallization of $In_2O_3$. (c) GIXRD spectra of $HfO_2/In_2O_3/HfO_2$ stack, $HfO_2/In_2O_3$ stack, and single $HfO_2$ layer together with standard cubic $In_2O_3$ and monoclinic $HfO_2$ patterns. EBSD images of (d) $HfO_2/In_2O_3/HfO_2$ stack and (e) $HfO_2/In_2O_3$ stack. (f) Distributions of grain size extracted from (d) and (e). The average grain size in $HfO_2/In_2O_3/HfO_2$ stack is nearly twice that of $HfO_2/In_2O_3$ stack.

**Acknowledgements**

This work was supported by National Key R&D Program of China under Grant 2022YFB3606900, the National Natural Science Foundation of China under Grant 62274107 and 92264204, and Shanghai Pilot Program for Basic Research-Shanghai Jiao Tong University under Grant 21TQ1400212.

**Author Contributions**

C.W. conceived the idea for dielectric deposition enhanced crystallization of the $In_2O_3$ channel. C.W., K.J., J.Z., Z.W., G.W., C.Z., and M.S. conducted all the data analysis. C.W. and M.S. co-wrote the manuscript and all authors commented on it.

**Competing interests**

The authors declare no competing interests.

**Additional information**

Additional details for structure and fabrication process of $In_2O_3$ transistors and Hall bar devices, $I_D$-$V_{DS}$ characteristics of BG $In_2O_3$ transistor with top $HfO_2$, CV characteristics of Mo/$HfO_2$/Mo stack, comparation between $\mu_H$, experimental $\mu_{FE}$, and calculated $\mu_{FE}$, extra HRTEM cross-section images of $HfO_2$/$In_2O_3$/$HfO_2$ stack and $HfO_2$/$In_2O_3$ stack and the derivation of relationship between $\mu_{FE}$ and intrinsic $\mu$ are in the supplementary information.

## Supplementary Information

# Dielectric Deposition Enhanced Crystallization in Atomic-Layer-Deposited Indium Oxide Transistors Achieving High Gated-Hall Mobility Exceeding 100 $cm^2/V\cdot s$ at Room Temperature

*Chen Wang[1], Kai Jiang[1], Jinxiu Zhao[1], Ziheng Wang[1], Guilei Wang[2], Chao Zhao[2], and Mengwei Si[1,*]*

[1]National Key Laboratory of Advanced Micro and Nano Manufacture Technology and School of Information Science and Electronic Engineering, Shanghai Jiao Tong University, Shanghai 200240, China;

[2]Beijing Superstring Academy of Memory Technology, Beijing 100176, China.

*Correspondence authors. Email: mengwei.si@sjtu.edu.cn

## 1. Structure and fabrication process of $In_2O_3$ transistors and Hall bar devices

Three different device structures including bottom-gate without top $HfO_2$ dielectric deposition (BG w/o top $HfO_2$), bottom-gate with top $HfO_2$ dielectric deposition (BG w/ top $HfO_2$) and top-gate (TG) are fabricated, as shown in Fig. S1.

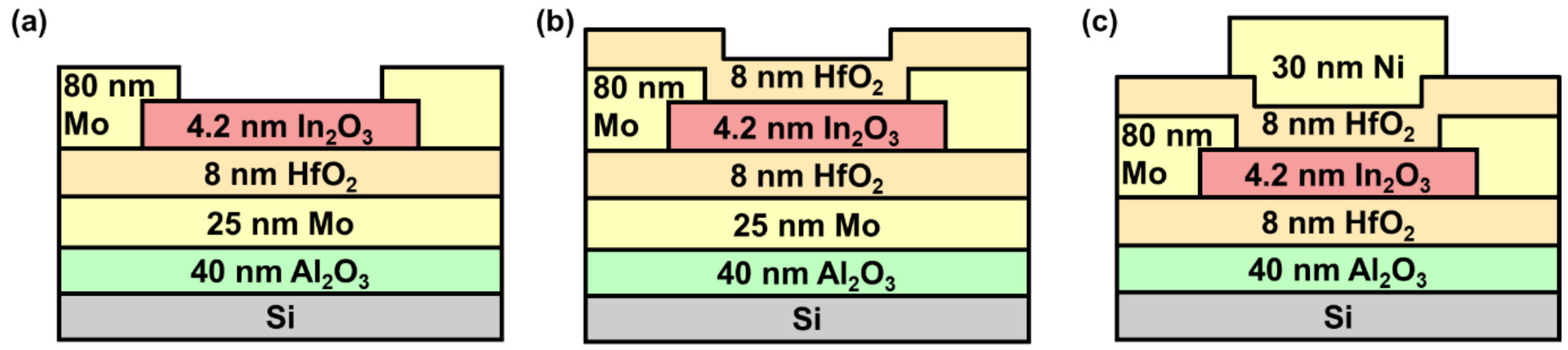

Figure S1. Schematic diagrams of ALD $In_2O_3$ transistors with various structures: (a) BG w/o top $HfO_2$, (b) BG w/ top $HfO_2$, (c) TG.

The fabrication process flow for the DG transistor as shown in Fig. S2, and other structures are fabricated with identical parameters together with certain processes omitted.

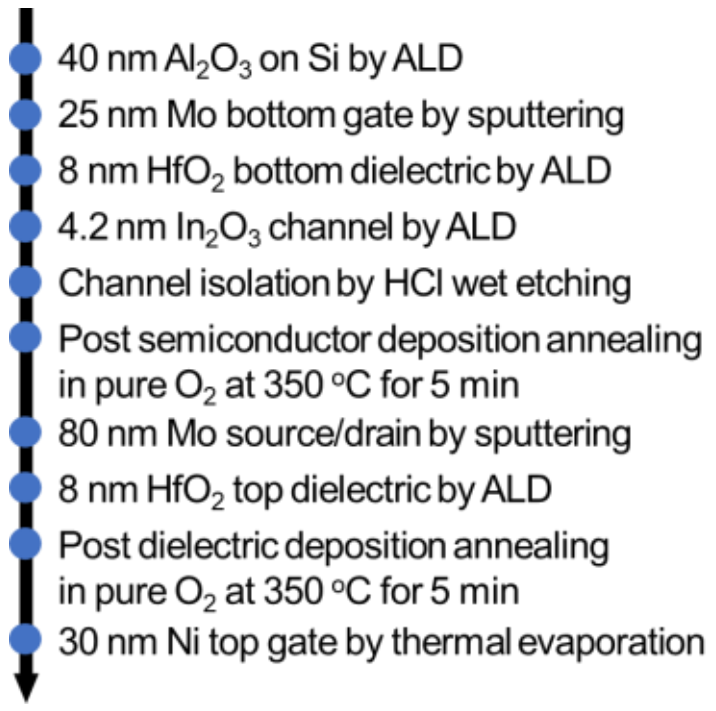

Figure S2. Illustration of the fabrication process of the DG $In_2O_3$ transistor. Other structures are fabricated by identical parameters with certain processes omitted.

## 2. $I_D$-$V_{DS}$ characteristics of BG $In_2O_3$ transistor with top $HfO_2$

The corresponding $I_D$-$V_{DS}$ characteristics of the same BG $In_2O_3$ transistor with top $HfO_2$ are plotted in Fig. S3, exhibiting a high maximum $I_D$ of 109 μA/μm and well-behaved drain current saturation at high $V_{DS}$.

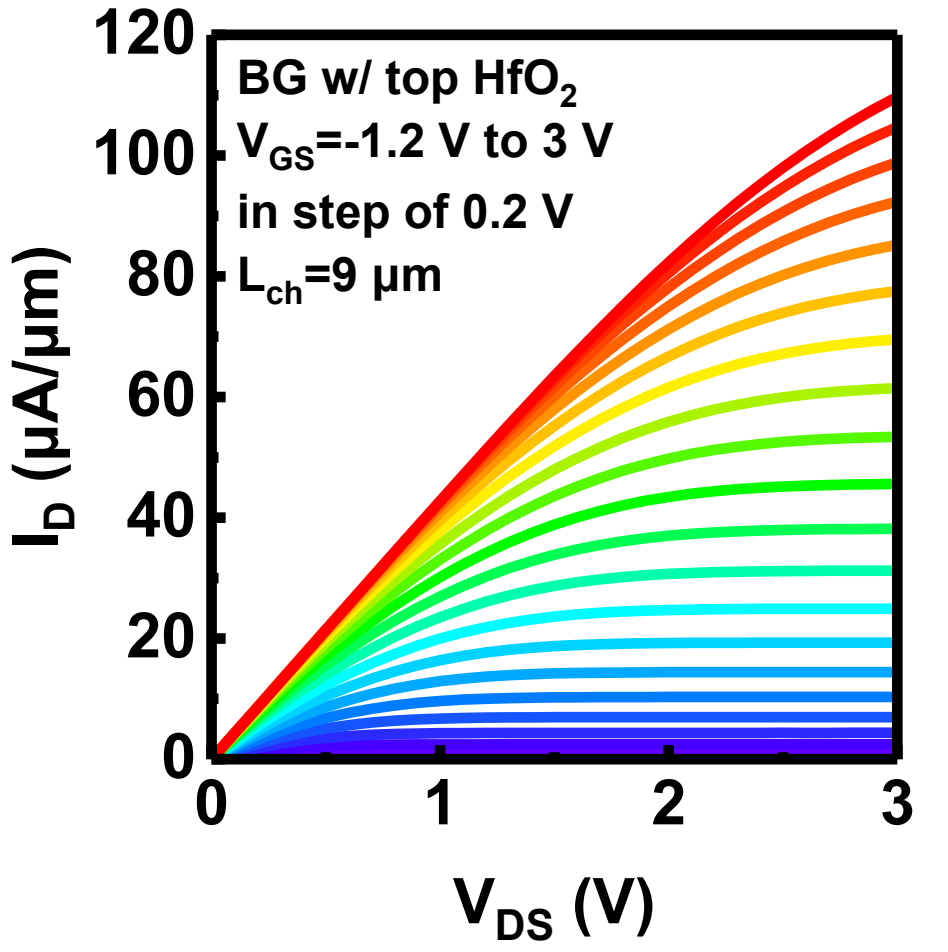

Figure S3. $I_D$-$V_{DS}$ characteristics of BG $In_2O_3$ transistor with top $HfO_2$ and with $L_{ch}$ of 9 μm.

### 3. $\mu_{FE}$ versus $V_{GS}$ of $In_2O_3$ transistors

$\mu_{FE}$ versus $V_{GS}$ characteristics of $In_2O_3$ transistors with various structures are plotted in Fig. S4. $\mu_{FE}$ of BG $In_2O_3$ transistors with top $HfO_2$ reaches 124.8 $cm^2/V \cdot s$ at 300 K and 137.9 $cm^2/V \cdot s$ at 100 K, which is much higher than that of BG $In_2O_3$ transistors without top $HfO_2$. However, $\mu_{FE}$ cannot be accurately extracted from $I_D$-$V_{GS}$ characteristics of TG and DG transistors due to the existence of hysteresis, as shown in Figs. S4(c) and S4(d).

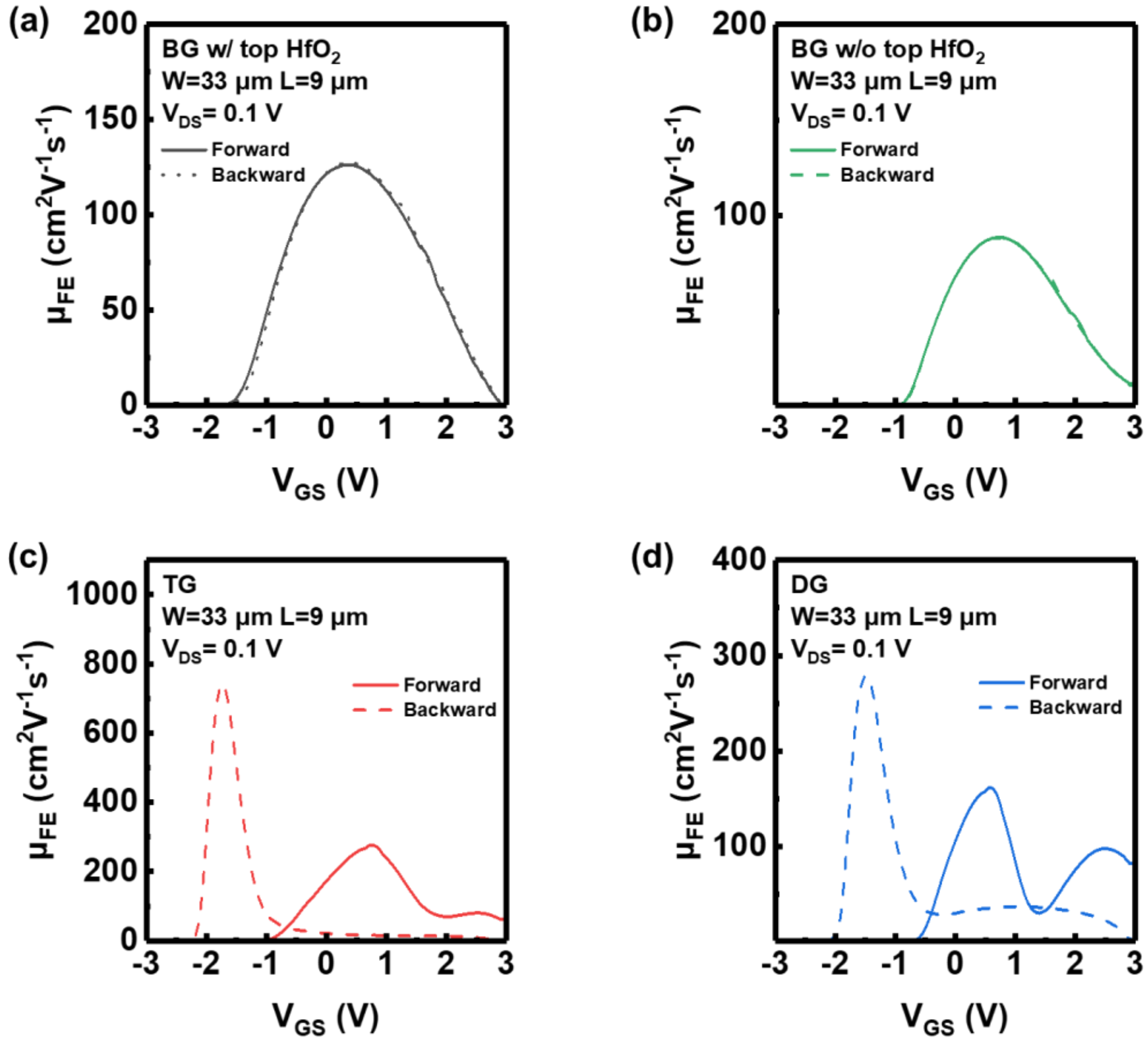

Figure S4. $\mu_{FE}$ versus $V_{GS}$ extracted from the transfer curves of $In_2O_3$ transistors with (a) BG with top $HfO_2$, (b) BG without top $HfO_2$, (c) TG, and (d) DG structures.

## 4. Capacitance-voltage (CV) characteristics

CV characteristics of Mo/$HfO_2$/Mo stack is plotted in Fig. S5. The $C_{OX}$ is determined to be 1.31 μF/cm$^2$ with frequency of 1 kHz and voltage of 0 V.

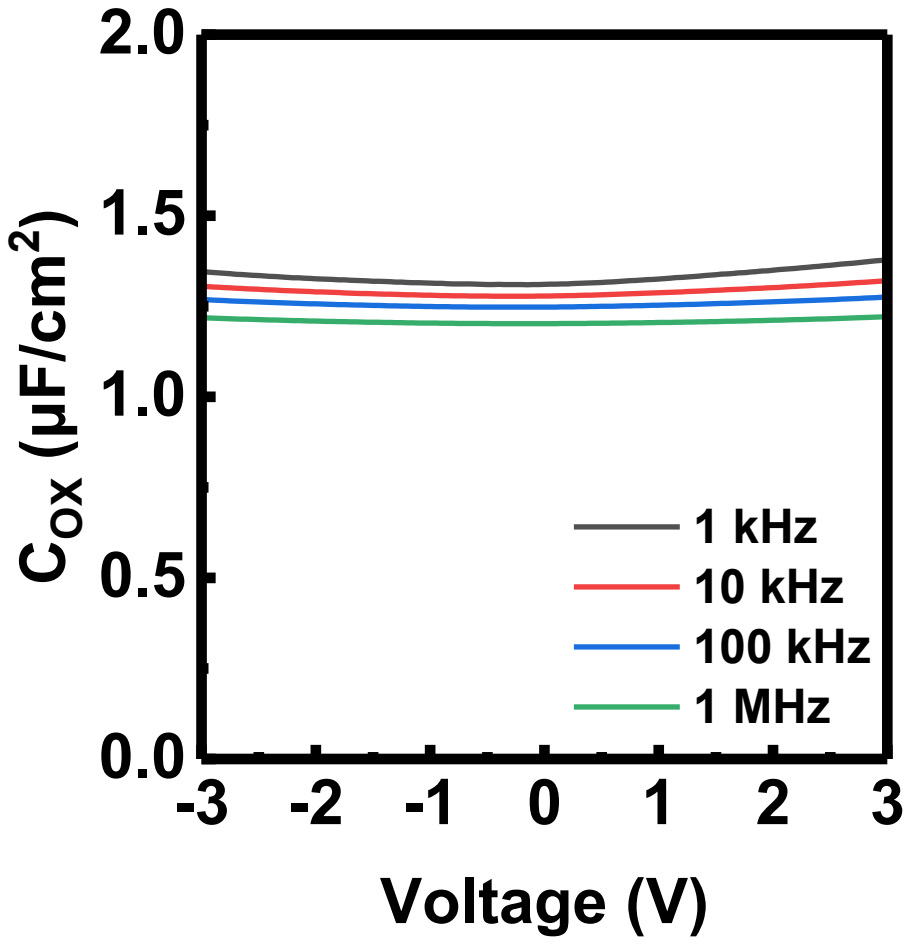

Figure S5. CV characteristics of Mo/$HfO_2$/Mo stack with frequency varying from 1 kHz to 1 MHz.

## 5. Comparation between $\mu_H$, experimental $\mu_{FE}$, and calculated $\mu_{FE}$

Fig. S6 compares $\mu_H$, experimental $\mu_{FE}$, and calculated $\mu_{FE}$ from $\mu_H$ by eqn. (1) of a BG $In_2O_3$ Hall bar device with top $HfO_2$ at 300 K. The calculated $\mu_{FE}$ from $\mu_H$ matches well with experimental $\mu_{FE}$.

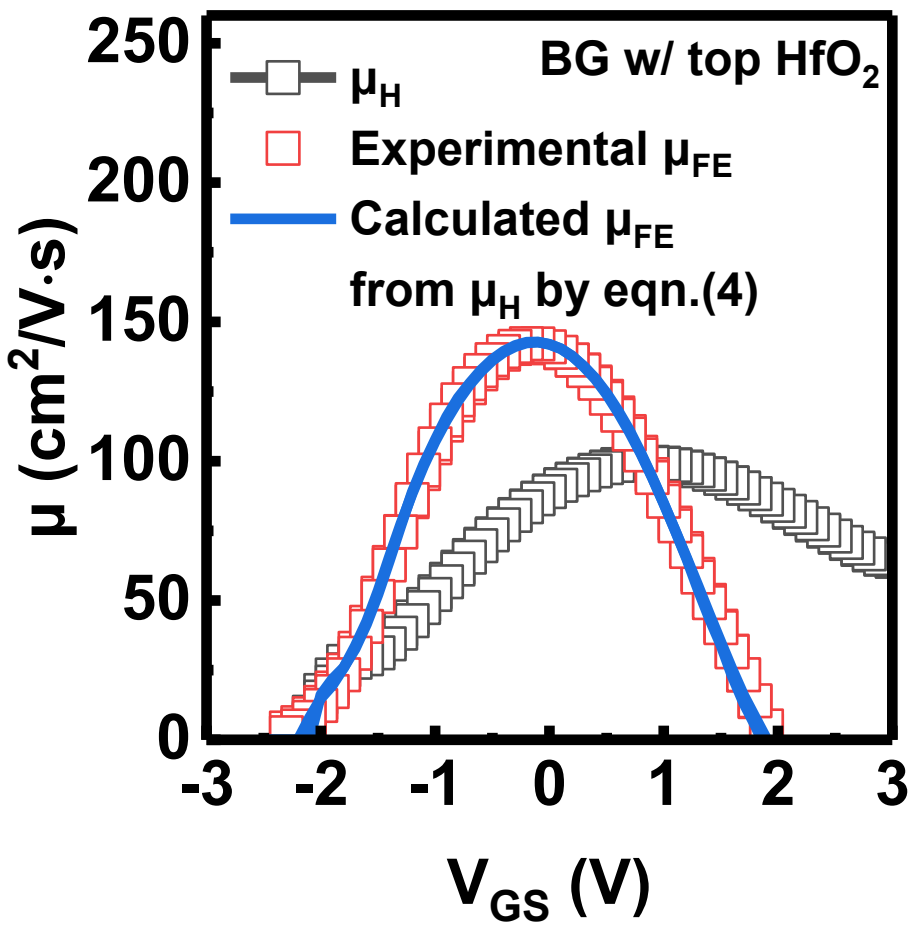

Figure S6. $\mu_H$, experimental $\mu_{FE}$, and calculated $\mu_{FE}$ from $\mu_H$ by eqn. (3) of BG $In_2O_3$ Hall bar device with top $HfO_2$ at 300 K.

## 6. HRTEM cross-section images of $HfO_2/In_2O_3/HfO_2$ stack and $HfO_2/In_2O_3$ stack

Additional HRTEM mages of $HfO_2/In_2O_3/HfO_2$ stack and $HfO_2/In_2O_3$ stack are given Fig. S7.

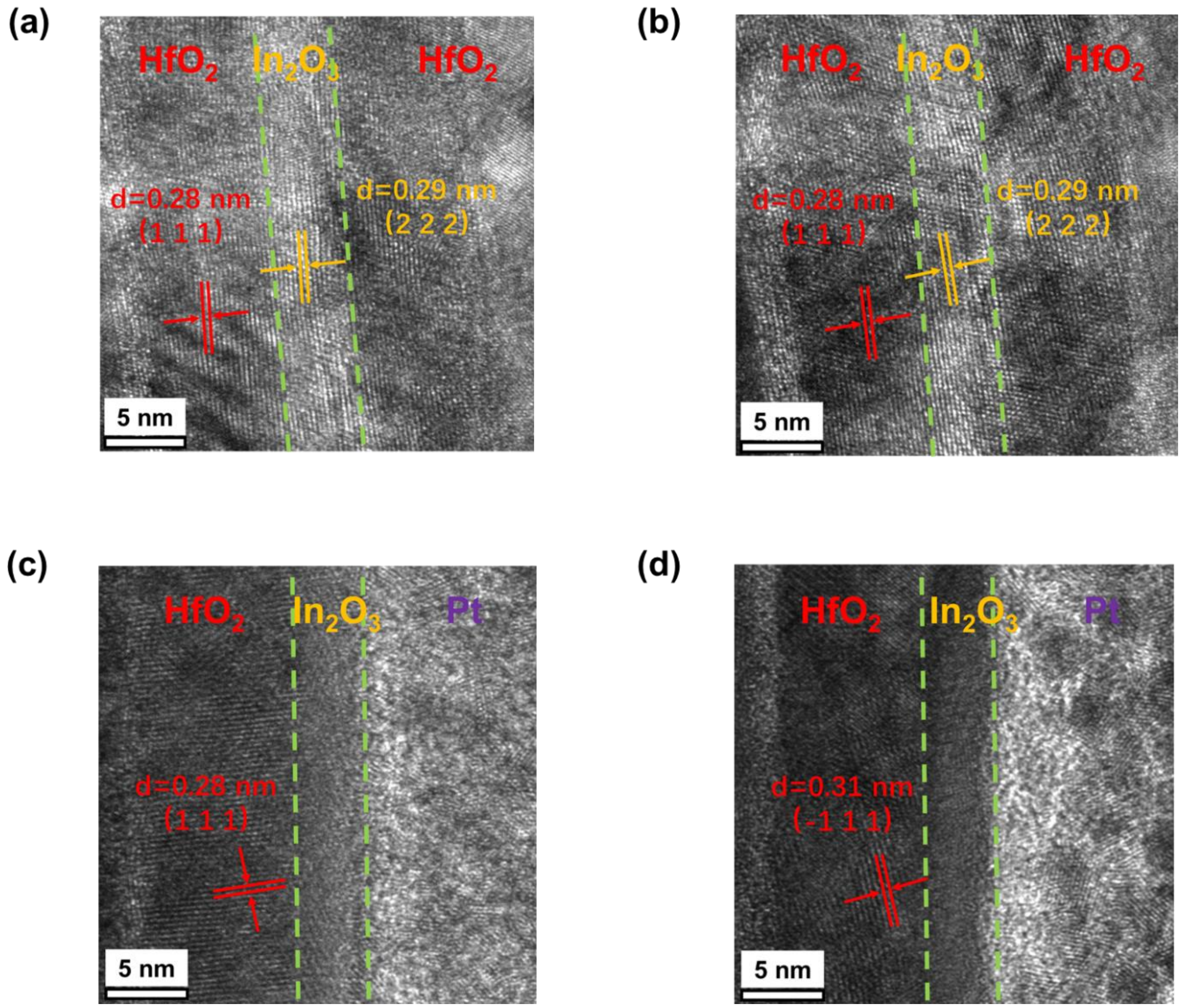

Figure S7. HRTEM cross-section images of (a), (b) $HfO_2/In_2O_3/HfO_2$ stack and (b), (d) $HfO_2/In_2O_3$ stack.

## 7. The derivation of relationship between $\mu_{FE}$ and intrinsic μ.

The $I_D$ of transistors is determined by[1]

$$I_D = \frac{W}{L}\mu Q_n V_{DS} = \frac{W}{L}\mu C_{ox}\left(V_{GS} - V_{TH} - \frac{V_{DS}}{2}\right)V_{DS} \quad \text{(S1)}$$

where $Q_n$ is the mobile carrier density, $C_{ox}$ is the gate capacitance, $V_{TH}$ is the threshold voltage, and $V_{DS}$ is the drain-to-source voltage. $\mu_{FE}$ is defined using $g_m$[1]

$$\mu_{FE} = \frac{L}{WC_{ox}V_{DS}}\left(\frac{\partial I_D}{\partial V_{GS}}\right) = \frac{g_m L}{WC_{ox}V_{DS}} \quad \text{(S2)}$$

Combining eqn. (1) and (2), the relationship between $\mu_{FE}$ and intrinsic μ is achieved as[2]

$$\mu_{FE} = \frac{\partial \mu}{\partial V_{GS}}\left(V_{GS} - V_{TH} - \frac{V_{DS}}{2}\right) + \mu \quad \text{(S3)}$$

**Reference:**

(1) Neamen, D. A. *Semiconductor Physics and Devices Basic Principles*; McGraw-Hill, 2012.

(2) Wang, C. *et al.* On the Accurate Evaluation of Intrinsic Electron Mobility on Oxide Semiconductor Transistors. *IEEE Electron Device Lett.* 2025.